# Conduction Electron Scattering and Spin-Flipping at Sputtered Co/Ni Interfaces.

H.Y.T. Nguyen, R. Acharyya, E. Huey, B. Richard, R. Loloee, W.P. Pratt Jr., and J. Bass
Department of Physics and Astronomy, Michigan State University, East Lansing, MI 48824.
Shuai Wang and Ke Xia
Department of Physics, Beijing Normal University, Beijing 100875, China

Current-perpendicular-to-plane magnetoresistance measurements let us quantify conduction electron scattering and spin-flipping at a sputtered ferromagnetic/ferromagnetic (F1/F2 = Co/Ni) interface, with important consequences for CPP-MR and spin-torque experiments with perpendicular anisotropy. We use ferromagnetically coupled [Ni/Co]$_n$Ni multilayers, and Py-based, symmetric double exchange-biased spin-valves (DEBSVs) containing inserts of ferromagnetically coupled [Co/Ni]$_n$Co or [Ni/Co]$_n$Ni multilayers, to derive Co/Ni interface specific resistances $AR^{\uparrow}_{Co/Ni} = 0.03^{+0.02}_{-0.03}$ f$\Omega$m$^2$ and $AR^{\downarrow}_{Co/Ni} = 1.00 \pm 0.07$ f$\Omega$m$^2$, and interface spin-flipping parameter $\delta_{Co/Ni} = 0.35 \pm 0.05$. The specific resistances are consistent with our no-free-parameter calculations for an interface thickness between 2 and 4 monolayers (ML) that is compatible with expectations.

From measurements of current-perpendicular-to-plane (CPP) magnetoresistance (MR) on sputtered multilayers, we now have good experimental information about scattering and spin-flipping of conduction electrons at non-magnetic/non-magnetic (N1/N2) interfaces [1,2], and scattering at ferromagnetic/non-magnetic (F/N) interfaces [1,2]. We also know a little about spin-flipping at F/N interfaces [3]. However, no-one has yet measured conduction electron scattering and spin-flipping at any F1/F2 interface. In this paper, we study such scattering and spin-flipping for Co/Ni. We chose Co and Ni because their lattice structures and lattice constants should be similar in sputtered multilayers, letting us calculate $AR^{\uparrow}_{Co/Ni}$ and $AR^{\downarrow}_{Co/Ni}$ with no adjustment, and their long bulk spin-diffusion lengths (see below) should let us isolate interfacial spin-flipping. Also, Co/Ni multilayers with very thin (≤ 1 nm) layers are widely used as the spin-polarizer in spin-torque studies with perpendicular magnetic anisotropy [4-6]. Since sputtered interfaces are typically a few monolayers (ML) thick [7], such multilayers are mostly interfaces. With the spin-flipping probability at Co/Ni interfaces that we find, only a few Co/Ni bilayers should 'saturate' spin-flipping, thereby limiting the growth of CPP-MR and spin-torque with increasing bilayer number.

When Co and Ni layers alternate, exchange interactions should cause their moments to align parallel to each other. Then, so long as the Co and Ni layers are thick enough that the magnetization lies in the layer plane [8,9], the magnetization of a short-wide [Co/Ni]$_n$ multilayer ($n$ is the number of repeats) should reverse as a unit at our measuring temperature of 4.2K. Assuming diffusive transport through sputtered Co/Ni interfaces, we derive the two parameters for interfacial scattering [10,11], which can be written either as $AR^{\uparrow}_{Co/Ni}$ and $AR^{\downarrow}_{Co/Ni}$, or as $AR^{*}_{Co/Ni} = [AR^{\downarrow}_{Co/Ni} + AR^{\uparrow}_{Co/Ni}]/4$ and $\gamma_{Co/Ni} = [AR^{\downarrow}_{Co/Ni} - AR^{\uparrow}_{Co/Ni}]/[AR^{\downarrow}_{Co/Ni} + AR^{\uparrow}_{Co/Ni}]$. $AR^{\uparrow}_{Co/Ni}$ and $AR^{\downarrow}_{Co/Ni}$ are the specific resistances (area A through which the CPP current flows times interface resistance) when the conduction electron moment is parallel ($\uparrow$) or antiparallel ($\downarrow$) to the common direction of the moments of the Co and Ni through which the electron is passing. $AR^{*}_{Co/Ni}$ and $\gamma_{Co/Ni}$ appear in the usual two-current series resistor (2CSR) [10,11] or Valet-Fert (VF) [11] models. The present paper focuses upon $AR^{\uparrow}_{Co/Ni}$ and $AR^{\downarrow}_{Co/Ni}$, which relate most directly to our calculations.

In addition to $AR^{\uparrow}_{Co/Ni}$ and $AR^{\downarrow}_{Co/Ni}$, we also derive the parameter for interfacial spin-flipping, $\delta_{Co/Ni}$. When spin-flipping occurs, VF analysis involves exponential-like functions with dimensionless arguments [11,12]. For spin-flipping in thickness t of a metal with spin-diffusion length $l_{sf}$, the argument is the ratio t/$l_{sf}$. For spin-flipping at a Co/Ni interface, the argument is $\delta_{Co/Ni}$. $\delta_{Co/Ni}$ is related to the probability $\mathcal{P}$ of spin-flipping at the interface by $\mathcal{P} = [1 - \exp(-\delta_{Co/Ni})]$.

To constrain our parameters we use three sets of samples. The first set is a simple [Ni(3)/Co(3)]$_n$Ni(3) multilayer, where we measure the total specific resistance, $AR_T$, vs $n$. Such data let us estimate $AR^{\uparrow}_{Co/Ni}$. To estimate $AR^{\downarrow}_{Co/Ni}$ and $\delta_{Co/Ni}$, we need a CPP-MR, which requires a ferromagnet with a different coercive field than that of the Co/Ni multilayer. We chose pinned Permalloy = Py ≈ Ni$_{0.8}$Fe$_{0.2}$. To enhance the CPP-MR's sensitivity to $\delta_{Co/Ni}$, we use a symmetric, double exchange-biased spin-valve (DEBSV) of the form: FeMn(8)/Py(6)/Cu(10)/X$_i$/Cu(10)/Py(6)/FeMn(8), where thicknesses are in nm, and $t_{Cu}$ = 10 nm is thick enough to exchange-decouple X$_i$ from the Py layers. Here the antiferromagnetic FeMn layers exchange-bias pin the two equal thickness Py layers so that their magnetic moments reverse together at a much higher field H than is needed to reverse the moment of X$_i$. X$_i$ is chosen as X$_1$ = [Co/Ni]$_n$Co or X$_2$ = [Ni/Co]$_n$Ni, with Co and Ni thicknesses of 3 nm, thick enough to give an in-plane moment, but thin enough so that spin-flipping within the Co (spin-diffusion length =

$l_{sf}^{Co} \sim 60$ nm [13]) and Ni ($l_{sf}^{Ni} \sim 21$ nm [14]) is weak enough to not mask the effects of δ. The CPP-MR then results from reversal of the moment of $X_i$ from parallel (P) to anti-parallel (AP) to the common direction of the moments of the two Py layers. We focus upon how AΔR = AR(AP) – AR(P) increases with *n,* an increase that depends sensitively upon $δ_{Co/Ni}$. By studying DEBSVs with both $X_1$ and $X_2$, we over-constrain the parameters of interest. Mirror symmetry of the samples simplifies the numerical calculations using VF theory [11], which need to be performed for only half of each sample.

Our sputtering system, sample preparation, and measuring techniques are described in ref. [15]. The multilayer samples were sputtered in a six gun, ultra-high vacuum compatible system, with the substrates held at temperatures between 243K and 303K. For the simple multilayers, all four metals (Nb, Cu, Co, and Ni) were sputtered from 2.25" diam. targets at rates ~ 0.4-0.6 nm/sec. For the more complex Py-based DEBSVs, the four large targets were used for Nb, Py, Co, and Ni, but Cu and FeMn were sputtered from 1" diam. targets with sputtering rates ~ 0.05-0.2 nm/sec. To achieve uniform current flow in the CPP geometry, the multilayers were sandwiched between ~ 1.1 mm wide, 150 nm thick, crossed Nb strips, that superconduct at our measuring temperature of 4.2K. We find the overlap area A ~ 1.2 mm² through which the CPP current flows by measuring the width of each Nb strip with a Dektak profilometer.

To analyze our data, we separately establish all of the other CPP-MR parameters of Co and Ni, and of the other constituents of the sample (FeMn, Py, and the various interfaces), thus leaving to be found from our new samples only the three parameters for Co/Ni interfaces: $AR_{Co/Ni}^\uparrow$, $AR_{Co/Ni}^\downarrow$, and $δ_{Co/Ni}$. To ensure that these other parameters are reliable, they are taken from measurements in our own laboratory with sputtering gas pressure, sputtering rates, etc., the same as those used for the present samples. Past experience, combined with regular rechecks, indicate that our derived parameters should be stable to within the uncertainties that we specify below. For examples of agreements of some of our previously derived parameters with no-free-parameter calculations see refs. [1,16] for interface specific resistances of closely lattice matched metal pairs, and refs. [3,17] for the scattering asymmetry and spin-diffusion length of Py.

The simple multilayers had the form: Nb(150)/Cu(5)/[Ni(3)/Co(3)]$_n$Ni(3)/Cu(5)/Nb(150). The DEBSVs had the form: Nb(150)/Cu(5)/FeMn(8)/Py(6)/Cu($t_{Cu}$=10)/$X_i$/Cu($t_{Cu}$=10)/Py(6)/FeMn(8)/Cu(5)/Nb(150), with $X_i$ given by either $X_1$ or $X_2$. We exchange-bias pinned the two outer Py layer magnetizations by holding the DEBSVs at 453K for 2 min in an in-plane magnetic field H ~ 200 Oe, and then cooling to room temperature in the field.

Table I. $AR_{Co/Ni}^\uparrow$ and $AR_{Co/Ni}^\downarrow$ for perfect interfaces for three different lattice constants.

| Lattice constants | $AR_{Co/Ni}^\uparrow$ (fΩm²) | $AR_{Co/Ni}^\downarrow$ (fΩm²) |
|---|---|---|
| Co (0.3549 nm) | 0.015 | 0.725 |
| Ni (0.3524 nm) | 0.024 | 0.728 |
| (Co+Ni)/2 (nm) | 0.019 | 0.731 |

Table II. $AR_{Co/Ni}^\uparrow$ and $AR_{Co/Ni}^\downarrow$ for three different interface structures.

| Interface Structure | $AR_{Co/Ni}^\uparrow$ (fΩm²) | $AR_{Co/Ni}^\downarrow$ (fΩm²) |
|---|---|---|
| Perfect | 0.015 | 0.725 |
| 2ML (50%-50%) | 0.016 | 0.864 |
| 4ML (50%-50%) | 0.018 | 1.193 |

To compare with our data, we calculated the interface specific resistances $AR_{Co/Ni}^\uparrow$ and $AR_{Co/Ni}^\downarrow$ for parallel ordering of the Co and Ni magnetizations using the no-free-parameter procedures described in refs. [16,18,19]. As the lattice parameters of Co and Ni are very close, we assumed a single fcc lattice for the whole sample. As usual for sputtering, the lattice normal was taken as <111>, perpendicular to the close-packed planes. From density functional theory, we first calculated the electronic structures of Co and Ni, using only assumed lattice constants and numbers of electrons. Table I shows that the (↑) and (↓) interface specific resistances for perfect interfaces are not sensitive to the lattice constant. For the rest of our calculations we use the constant for Co.

Having calculated the electronic structures for the Co lattice constant, we used the appropriate Landauer formula (corrected for the Sharvin resistance) [18] to calculate $AR_{Co/Ni}^\uparrow$ and $AR_{Co/Ni}^\downarrow$ for three different interface structures: (a) perfect interfaces; (b) 2 monolayers (ML); and (c) 4 ML of a randomly disordered 50%-50% interfacial alloy. The values in Table II will be compared with our experimental results. The uncertainties in the calculations are typically a few percent, due mostly to uncertainties in the calculated Fermi energies [16,19]. The most important predictions from Tables I and II are that $AR_{Co/Ni}^\uparrow$ is near zero, and that $AR_{Co/Ni}^\downarrow \gg AR_{Co/Ni}^\uparrow$.

For the numerical analysis of our complex multilayers with VF theory, we need to know the parameters for all of their layers and interfaces other than those that we wish to determine. Our own prior studies give the following parameters [3,14,20,21]: $ρ_{FeMn}$ = 875 ± 50 nΩm; $AR_{Nb/FeMn}$ = 1.0 ±0.6 fΩm²; $AR_{FeMn/Py}$ = 1.0 ± 0.4 fΩm²; $ρ_{Py}$ = 123 ± 40 nΩm; $β_{Py}$ = 0.76 ± 0.07; $l_{sf}^{Py}$ = 5.5 ± 1 nm; $AR_{Py/Cu}^*$ = 0.50 ± 0.04 fΩm²; $γ_{Py/Cu}$ = 0.7 ± 0.1. $ρ_{Co}$ = 60 ± 4 nΩn; $β_{Co}$ = 0.46 ± 0.05; $γ_{Co/Cu}$ = 0.75 ± 0.04; $AR_{Co/Cu}^*$ = 0.52 ± 0.02; $l_{sf}^{Co}$ = 60 ± 20 nm [13,22]; $ρ_{Ni}$ = 33 ± 3 nΩm; $β_{Ni}$ = 0.14 ± 0.02; $l_{sf}^{Ni}$ = 21 ± 2 nm; $γ_{Ni/Cu}$ = 0.29 ± 0.05; $AR_{Ni/Cu}^*$ = 0.18 $\pm 0.03$ f$\Omega$m$^2$; $\delta_{Co/Cu} = 0.33^{+0.03}_{-0.08}$. For Nb/Ni, ref. [20] plus new measurements give $AR_{Nb/Ni} = 2.5 \pm 0.5$. For Cu we only need the resistivity for the small target, which we found from Van der Pauw (VdP) measurements on separate 200 nm thick films to be $\rho_{Cu} = 10 \pm 1$ f$\Omega$m$^2$. From ref. [2] we estimate $l_{sf}^{Cu} \geq 1000$ nm. We checked that this is long enough to give results indistinguishable from those for $l_{sf}^{Cu} = \infty$, which we use to simplify the analysis. These parameters leave as unknowns only $AR_{Co/Ni}^{\uparrow}$, $AR_{Co/Ni}^{\downarrow}$, and $\delta_{Co/Ni}$.

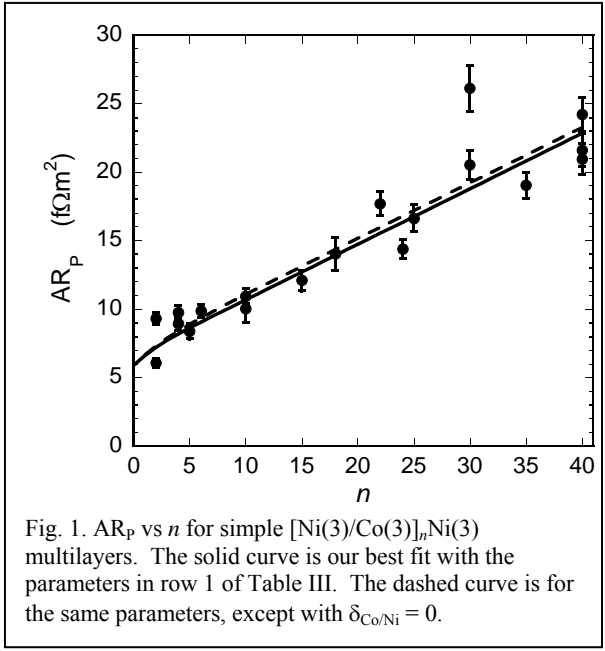

Fig. 1. $AR_P$ vs $n$ for simple $[Ni(3)/Co(3)]_n Ni(3)$ multilayers. The solid curve is our best fit with the parameters in row 1 of Table III. The dashed curve is for the same parameters, except with $\delta_{Co/Ni} = 0$.

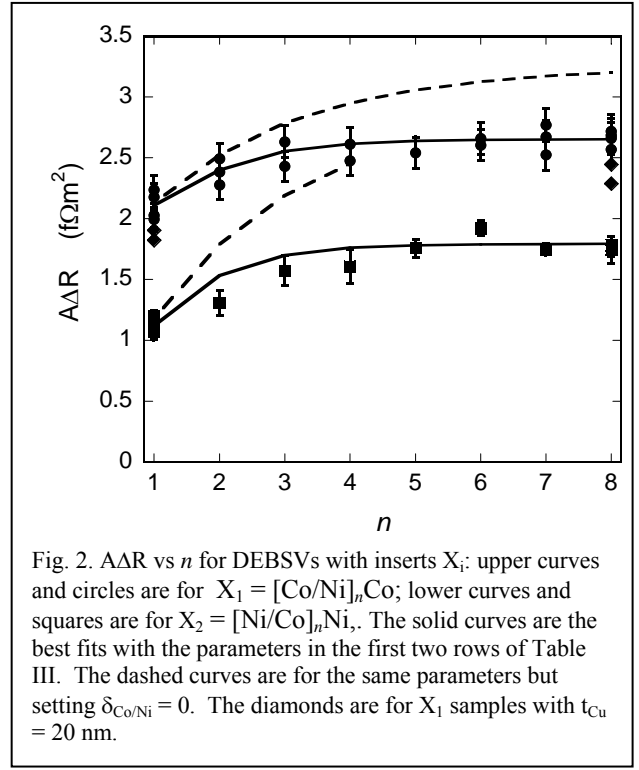

Fig. 2. $A\Delta R$ vs $n$ for DEBSVs with inserts $X_i$: upper curves and circles are for $X_1 = [Co/Ni]_n Co$; lower curves and squares are for $X_2 = [Ni/Co]_n Ni$. The solid curves are the best fits with the parameters in the first two rows of Table III. The dashed curves are for the same parameters but setting $\delta_{Co/Ni} = 0$. The diamonds are for $X_1$ samples with $t_{Cu} = 20$ nm.

We analyse both our multilayer and DEBSV data numerically using VF theory, including spin-flipping within Co, Ni, Py, and at the Co/Cu and Co/Ni interfaces. For these complex samples, we match boundary conditions at layer boundaries and solve the resulting linear equations by computer. Examples of VF equations for simpler multilayers involving spin-flipping at interfaces are given in ref. [12]. Our analysis involves an iterative procedure with the following three sets of data. (1) The $AR_P$ data of Fig. 1. These data are insensitive to $\delta_{Co/Ni}$. (2) $A\Delta R$ ($n = 1$) of Fig. 2. This value is insensitive to $\delta_{Co/Ni}$. (3) $A\Delta R$ for $n = 7$, which we choose as the average of the data for $n = 6$-$8$. This value is highly sensitive to $\delta_{Co/Ni}$. The analysis gives the Co/Ni parameters listed in Table III. To give the reader a feel for which unknown parameters dominate each set of samples, we use 2CSR simplifications.

In the simple $[Ni(3)/Co(3)]_n Ni(3)$ multilayer, for large $n$, the low resistance $\uparrow$ channel nearly 'shorts' the sample. For full shorting, the 2CSR model would give:

$$AR_T = 4AR_{Nb/Ni} + \rho_{Ni}^{\uparrow}(3) + n[\rho_{Co}^{\uparrow}(3) + \rho_{Ni}^{\uparrow}(3) + 2AR_{Co/Ni}^{\uparrow}]. \quad (1)$$

Our main interest is in the slope of $AR_T$ vs $n$ for large $n$. From the values given above, the sum of the first two terms in the slope should be $\approx 0.4$ f$\Omega$m$^2$. Since, from Table II we expect $2AR_{Co/Ni}^{\uparrow}$ to be an order of magnitude smaller than this value, the uncertainty in $2AR_{Co/Ni}^{\uparrow}$ will be comparable to its value. Our fit to the scattered data depends also upon the value of $4AR_{Nb/Ni}$.

For the symmetric DEBSVs and $n = 1$, $A\Delta R$ should be insensitive to $\delta_{Co/Ni}$ and well described by a 2CSR model. For $X_1$, the important term is proportional to:

$$(3)(\rho_{Ni}^{\downarrow} - \rho_{Ni}^{\uparrow}) + 2\times(3)(\rho_{Co}^{\downarrow} - \rho_{Co}^{\uparrow}) + 2(AR_{Co/Ni}^{\downarrow} - AR_{Co/Ni}^{\uparrow}) \quad (2).$$

From Table II, $2AR_{Co/Ni}^{\downarrow}$ now dominates, at about 75% of the total. So our estimate of $2AR_{Co/Ni}^{\downarrow}$ should be good to $\sim$ 10%. We test its reliability by independently fitting data for $X_1$ and $X_2$. The two cases differ only in an extra Co or Ni layer and two interfaces with either $2AR_{Co/Cu}$ or $2AR_{Ni/Cu}$.

To show the importance of $\delta_{Co/Ni}$, we include curves with and without its contribution.

Fig. 1 shows $AR_T$ vs $n$ for the simple $[Ni(3)/Co(3)]_n Ni(3)$ multilayers. The solid curve is our best fit with the parameters in Table III for $X_1$. The dashed curve for $\delta_{Co/Ni} = 0$ shows that the data in Fig. 1 are insensitive to $\delta_{Co/Ni}$. Due to the scatter in the data, and uncertainty in $AR_{Nb/Ni}$, we can say from these data alone only that $0 \leq AR_{Co/Ni}^{\uparrow} \leq 0.05$ f$\Omega$m$^2$. The DEBSV data are needed to find $AR_{Co/Ni}^{\uparrow}$ more precisely.

Fig. 2 shows A$\Delta$R vs $n$ for DEBSVs with: (a) circles for X$_1$ and (b) squares for X$_2$. The solid curves are the respective best fits with the parameters given in the top two rows of Table III. The dashed curves show the results if we set $\delta_{Co/Ni} = 0$—as expected, the data for $n \geq 2$ are sensitive to $\delta_{Co/Ni}$.

Table III. $AR_{Co/Ni}^{\uparrow}$ and $AR_{Co/Ni}^{\downarrow}$ (in f$\Omega$m$^2$), and $\delta_{Co/N}$, for sample sets X$_1$ and X$_2$ with t$_{Cu}$ = 10 nm, X$_1$ with t$_{Cu}$ = 20 nm, and our 'best estimates'.

| [F1/F2]$_n$/F1 (t$_{Cu}$) | $AR_{Co/Ni}^{\uparrow}$ | $AR_{Co/Ni}^{\downarrow}$ | $\delta_{Co/Ni}$ |
|---|---|---|---|
| X$_1$ = [Co/Ni]$_n$Co (10) | $0.03_{-0.03}^{+0.02}$ | 1.08±0.15 | 0.32±0.1 |
| X$_2$ = [Ni/Co]$_n$Ni (10) | $0.03_{-0.03}^{+0.02}$ | 1.00±0.07 | 0.38±0.05 |
| X$_1$ = [Co/Ni]$_n$Co (20) | $0.04_{-0.04}^{+0.02}$ | 0.73±0.15 | 0.24±0.1 |
| Weighted Averages | $0.03_{-0.03}^{+0.02}$ | 1.00±0.07 | 0.35±0.05 |

Table III shows that the data of Figs. 1 and 2 for t$_{Cu}$ = 10 nm (rows 1 and 2) are consistent with a single set of parameters to within uncertainties. For completeness, we note that data for additional DEBSVs with form X$_1$ but t$_{Cu}$ = 20 nm (diamonds in Fig. 2) lie about three times as far below the circles as we had expected, mainly due to an apparently smaller $AR_{Co/Ni}^{\downarrow}$ (row 3 of Table III). The reason for this deviation is under further investigation. We take as our best estimates, the weighted averages of the values in all three rows 1-3. These average values for $AR_{Co/Ni}^{\uparrow}$ and $AR_{Co/Ni}^{\downarrow}$ are consistent with the calculated values in Table II for a Co/Ni interface thickness between 2ML and 4 ML, compatible with expectation for interface intermixing [7].

To summarize, we have determined the parameters for conduction electron scattering and spin-flipping at sputtered F1/F2 = Co/Ni interfaces, from CPP-MR measurements on simple [Ni/Co]$_n$Ni multilayers, and two different forms of Py-based double exchange-biased spin-valves (DEBSVs) containing ferromagnetically coupled Co/Ni multilayers of either [Ni/Co]$_n$Ni or [Co/Ni]$_n$Co, bounded by 10 nm thick Cu layers, We found good agreement between parameters derived for the two sets of completely independent DEBSVs, and reasonable agreement for a few samples bounded by 20 nm thick Cu layers. These agreements support the reliability of our best estimates: $AR_{Co/Ni}^{\uparrow}$ = $0.03_{-0.03}^{+0.02}$ f$\Omega$m$^2$, $AR_{Co/Ni}^{\downarrow}$ = 1.00±0.07 f$\Omega$m$^2$, and $\delta_{Co/Ni}$ = 0.35±0.05, which are weighted averages of the parameters from all three sets of samples. These values of $AR_{Co/Ni}^{\uparrow}$ and $AR_{Co/Ni}^{\downarrow}$ are also consistent with our calculations for interface thicknesses between 2 ML and 4 ML, which is compatible with expectation [7]. We look forward to theoretical guidance as to whether this large $\delta_{Co/Ni}$ is due more to spin-orbit interactions, or to misaligned moments at the Co/Ni interface [23]. These parameters will allow one to calculate the contributions of a Co/Ni multilayer to CPP-MR or spin-torque circuits. They are particularly important for perpendicular anisotropy samples, where the Co and Ni layers are so thin that bulk effects are small. Also important is the limitation that the relatively large value of $\delta_{Co/Ni}$ sets upon the ability to increase A$\Delta$R or the spin-polarization by adding more Co/Ni interfaces.

Acknowledgments: This work was supported in part by US-NSF grant DMR 08-04126, a Korea Institute for Science and Technology (KIST) grant, and Chinese grants NSF:10634070 and MOST:2011CB921803.